\documentclass[journal]{vgtc}              

\usepackage{enumitem}
\usepackage[table,xcdraw,svgnames]{xcolor}
\usepackage{tikz}

\onlineid{0}

\vgtccategory{Research}

\vgtcpapertype{please specify}

\title{\grammar: A Declarative Grammar for Network Construction, Transformation, and Interactive Visualization}

\author{%
  \authororcid{James Scott-Brown}{0000-0001-5642-8346},
  \authororcid{Alexis Pister}{0000-0002-2817-020X}, and 
  \authororcid{Benjamin Bach}{0000-0002-9201-7744}
}

\authorfooter{
  \item
    James Scott Brown, Alexis Pister, Benjamin Bach are with the University of Edinburgh.
}

\abstract{%
This paper introduces \grammar, a domain-specific language and declarative grammar for interactive network visualization design that supports multivariate, temporal, and geographic networks.
\grammar{} allows users to specify network visualizations 
as combinations of
primitives and building blocks.
These support network creation and transformation, including computing metrics; orderings, seriations and layouts; visual encodings, including glyphs, faceting, and label visibility; and interaction for exploration and modifying styling.
This approach allows the creation of a range of visualizations including many types of node-link diagrams, adjacency matrices using diverse cell encodings and node orderings, arc diagrams, PivotGraph, small multiples, time-arcs, geographic map visualizations, and hybrid techniques such as NodeTrix.
\grammar{} aims to 
remove the need to use multiple
libraries for analysis, wrangling, and visualization. Consequently, \grammar{} supports the agile development of applications for visual exploration of networks and data-driven storytelling.
Documentation, source code, further examples, and an interactive online editor can be found online: \projectURL.
}


\teaser{
  \centering
  \includegraphics[width=\linewidth]{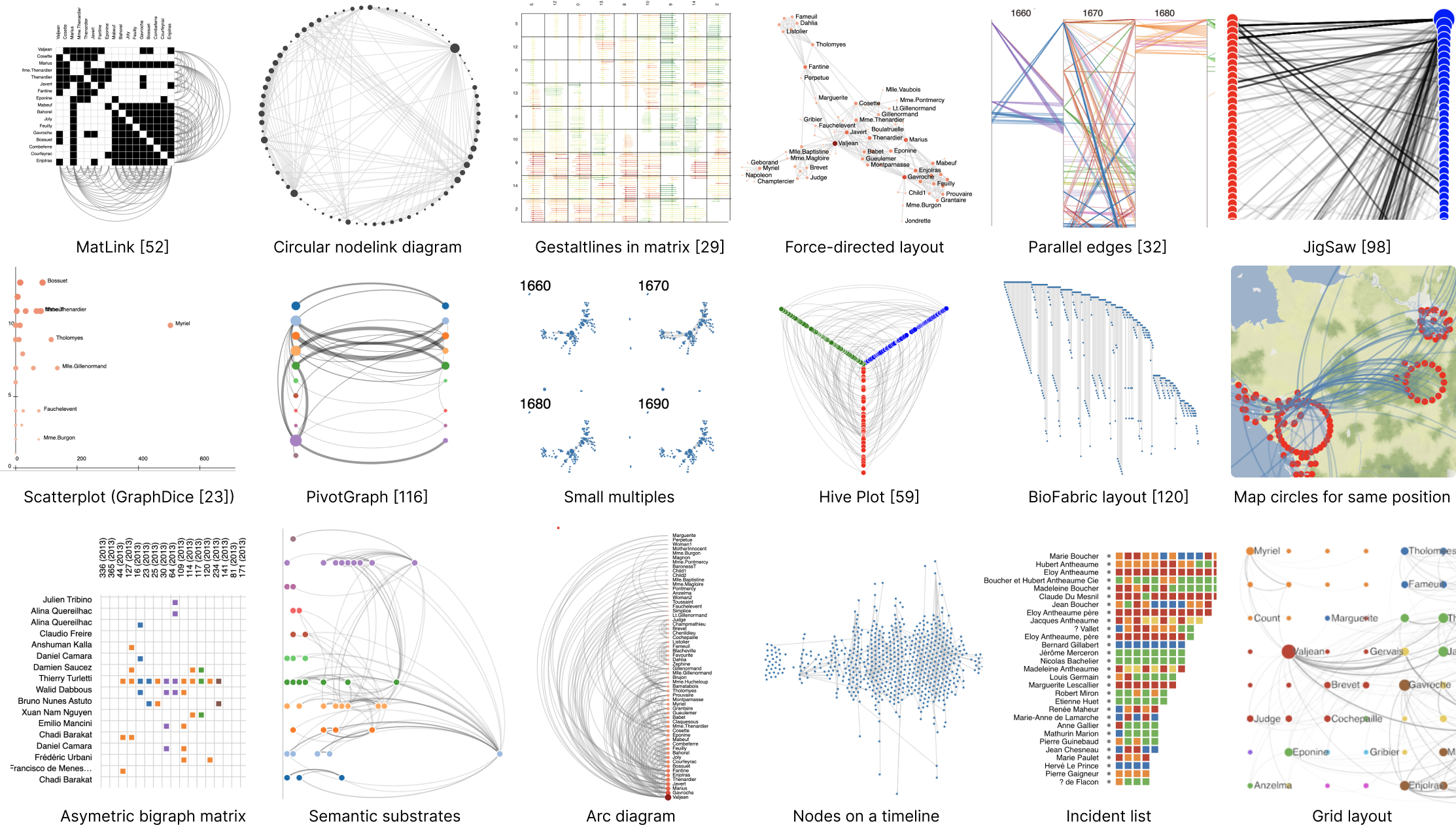}
  \caption{Network visualization designs created with \grammar.}
  \label{fig:teaser}
}

\graphicspath{{figs/}{figures/}{pictures/}{images/}{./}} %

\usepackage{tabu}
\usepackage{booktabs}
\usepackage{lipsum}
\usepackage{mwe}

\usepackage{mathptmx}                  
\usepackage{marginnote}

\newcommand{\grammar}{NetPanorama}
\newcounter{bencounter}
\usepackage{soul}

\renewcommand{\st}[1]{}

\newcommand{\codeown}[1]{\textcolor{orange}{\texttt{#1}}}
\newcommand{\codeowna}[2]{\textcolor{orange}{\texttt{#1}}:~\textcolor{teal}{\texttt{#2}}}
\newcommand{\codeownv}[1]{\textcolor{teal}{\texttt{#1}}}
\newcommand{\ca}[1]{\codeown{#1}}
\newcommand{\cv}[1]{\codeownv{#1}}

\newcommand{\green}[1]{\textcolor{green}{#1}}
\newcommand{\blue}[1]{\textcolor{blue}{#1}}

\newcommand{\projectURL}[0]{\url{https://netpanorama.netlify.app/}}

\usepackage{wrapfig}

\begin{document}

\maketitle
\section{Introduction}

Researchers and designers have created numerous visualization techniques to provide complementary lenses on complex multivariate, temporal, and geographic 
networks.
These techniques (summarized in surveys including ~\cite{STAR_dynamic,geospatial_networks,STAR_multilayer,STAR_multivariate,STAR_large,STAR_groups}) go far beyond node-link diagrams, and span an enormously rich design space that includes
adjacency matrices with diverse cell glyphs (e.g.,~\cite{henry2006matrixexplorer,brandes2003visualizing}), adjacency lists, time-arcs for temporal networks~\cite{TimeRadarTrees}, 
parallel node-link diagrams (JigSaw)~\cite{stasko2008jigsaw}, HivePlots~\cite{krzywinski2012hive},
scatterplots~\cite{bezerianos2010graphdice}, or hybrid techniques such as NodeTrix~\cite{henry2007nodetrix} or MatLink~\cite{henry2007matlink} (\autoref{fig:teaser}).

Leveraging the richness of these techniques in applications, however, 
is challenging.
In most cases, visualization design requires early prototyping and user testing, quick design iterations, mixing and matching ideas from existing designs, testing different data sets~\cite{walny2019data}, and ensuring visual consistency across multiple visualizations. 
When collaborators are unfamiliar with proposed visualization designs, it may be necessary to demonstrate them by applying them to their own data.
However, 
software implementations of specific techniques may be unavailable or unsuitable for integration into new projects, whilst general-purpose frameworks for network visualization (e.g.,~\cite{Cytoscapejs,sigmajs,ggnetwork}) lack many of the concepts and
functionality for network visualization design beyond the node-link diagram. 
Many network visualization techniques also rely on computing specific metrics or applying transformations to the network topology (e.g., clustering, aggregation, projections, faceting, filtering), and implementing visualizations is often time-consuming and requires specialist knowledge; the effort further increases for applications that involve multiple views or interactivity, and those requiring visualizations to be robust to different types and sizes of networks. In short, network visualization consists of many steps that need design, implementation, and often iteration.

With \grammar{}, we introduce a unified domain-specific language (DSL) in the form of a declarative grammar~\cite{mcnutt2022no} that can specify a wide range of network visualization techniques and their required transformations.
\grammar{} provides a set of novel primitives and routines (\textit{constructs}) describing networks and their semantics across multiple sources and tables, calculating network metrics and applying transformations to a network's topology, specifying diverse layouts and node seriations~\cite{reorderJS}, creating rich visual glyphs for nodes and links, and defining strategies for label visibility and interaction specific to network data. This approach is similar to \textit{graph level operations}~\cite{glo-stix} in that it allows fine-grained design decisions, however \grammar{} is more expressive and adopts a declarative programming paradigm. 

The challenge in creating a grammar for network visualization is to review, define, and formalize common mechanisms in visualization design, and to integrate these into a coherent framework. Our respective choices are informed by five design goals which in turn are derived from analyzing existing libraries and processes for network visualization (\autoref{sec:criteria}). Inspired by other visualization grammars~\cite{2017-vega-lite, mcnutt2022no}, \grammar{} specifications are written as JSON and interpreted by our reference implementation. Our reference implementation is partly based on Vega and other network visualization libraries (\autoref{sec:implementation}) and renders in a browser. However, unlike Vega, \grammar{} provides the missing constructs for network data within a common framework while exposing some functionality from those other libraries (layouts~\cite{WebCoLa,tulip5}, matrix seriation~\cite{reorderJS}, and visual marks~\cite{2017-vega-lite}).

After reviewing related work (\autoref{sec:related-work}), 
we describe the five design criteria that motivated the development of \grammar{} (\autoref{sec:criteria}), 
explain the constructs structuring the grammar and how they address these criteria
(\autoref{sec:concepts}).
After details on the implementation (\autoref{sec:implementation}), we demonstrate the expressiveness of \grammar{} through examples of a range of existing visualization techniques, 
and reflect on our experience using \grammar{} in recent network visualization projects
(\autoref{sec:examples}).
More examples, documentation, and an interactive editor are found online: \projectURL{}.

This paper makes several contributions:
\begin{enumerate}[noitemsep]
    \item It identifies and defines a set of \textbf{constructs and primitives} to specify many aspects of network visualizations from data import to interactivity.
    \item It integrates these into a \textbf{unified declarative grammar} that allows users to fully specify interactive network visualizations, without requiring knowledge of other libraries. 
    \item It comes with a wide set of \textbf{example specifications} for  existing network visualization techniques.
    \item It provides a \textbf{reference implementation} with \textbf{interactive online editor} that can render interactive network visualizations in the browser from these specifications.
\end{enumerate}     

\section{Related Work}
\label{sec:related-work}

We review tools and libraries for network visualization and wrangling, visualization grammars, and grammars specific to network visualization, and describe how \grammar{} fills the gap left by these approaches (see \autoref{tab:comparison} for a structured comparison to the most relevant ones).

\textbf{Graphical network visualization tools} require no programming knowledge as they provide graphical interfaces for designing and visualizing networks. However, examples such as Gephi~\cite{Gephi}, Cytoscape~\cite{cytoscape}, SoNIA~\cite{SoNIA},  NodeXL~\cite{NodeXL, nodeXL2}, Visone~\cite{baur2001visone}, NetworkWorkbench~\cite{NetworkWorkbench}, SYF~\cite{SYF}, Pajek~\cite{Pajek}, UCINET~\cite{UCINET}, and NetMiner~\cite{NetMiner} exclusively focus on node-link diagrams, primarily with force-directed layouts.
General purpose visualization tools, whether template-based (Rawgraphs~\cite{rawgraphs}, DataWrapper~\cite{DataWrapper}) or mark-based (DataIllustrator~\cite{data_illustrator}, Charticulator~\cite{charticulator}, Data-Driven Guides~\cite{kim2016data}), offer no direct functionality for networks, their data structures, metrics, and queries (e.g., common neighbors).
General purpose \textbf{visualization programming libraries} such as ProtoVis~\cite{protovis}, D3~\cite{D3}, or Processing~\cite{processing}, are also not specific to networks but provide an abstraction layer for data visualizations. Again, \textbf{libraries for network visualization} almost exclusively support node-link diagrams (albeit with diverse styling options):
NetworkX~\cite{networkX} and PyVis~\cite{pyvis}, Graph-tool~\cite{graph_tool}, Plotly, and Netwulf~\cite{netwulf} in the Python ecosystem, 
and ggnetwork~\cite{ggnetwork}, ggraph~\cite{ggraph}, Igraph~\cite{igraph}, NetworkD3, NDTV-D3, and visNetwork~\cite{visNetwork} in the \texttt{R} ecosystem. Some of these display the generated visualization in a web browser, with embedded controls to adjust the appearance. Dedicated web-based libraries for network visualizations include G6~\cite{G6}, Cytoscape.js~\cite{Cytoscapejs}, and Sigma.js~\cite{sigmajs}.  
\textbf{Hybrid approaches} combine a GUI and a programming language. For example, the graph exploration system GUESS~\cite{GUESS,tulip5} embeds an interactive Python interpreter within a GUI; users can adjust what is displayed by coding commands or directly interacting with the displayed visualization.

A common limitation of all these existing authoring tools is a lack of expressiveness for visualizing network data: they can tweak the appearance of a node-link diagram, but are limited in producing other types of visualizations (\autoref{fig:teaser}), perform analysis, and create networks.

Complementing visualization capabilities, \textbf{network wrangling and data manipulation} are crucial to many real-world applications, including the definition of network schemas for table data (determining which parts of the data are to be interpreted as nodes and what are the semantics of links).
When data come as one or more tables, a network must be defined by defining node and link semantics from the tables, and potentially cleaning and transforming these networks through filtering, aggregation, or projection. 
Existing graphical tools include Ploceus~\cite{ploceus}, Orion~\cite{orion}, and Origraph~\cite{origraph}. However, none of these tools provides a textual domain-specific language for network creation and manipulation, making the reuse and generation of routines hard. Orion can save constructed workflows as XML, but this is not well suited to editing or rapid prototyping. \grammar{} shares some of these operations but is distinguished by being a grammar for network visualization that includes network wrangling capabilities, rather than a network wrangling tool with visualization capabilities: it is designed to allow the flexible creation of a wide range of visualizations, rather than being restricted to predefined visualization templates.

\textbf{Visualization Grammars} provide a higher-level way for designers to define visualizations, hiding implementation complexity and reducing coding effort. Beginning with Wilkinson's seminal \textit{Grammar of Graphics}~\cite{wilkinson1999the}, declarative visualization grammars were popularized by ggplot2~\cite{Wickham2010}. Vega later provided a browser-based \textit{declarative} implementation of a visualization grammar with its interactive and higher-level versions, Reactive Vega~\cite{2016-reactive-vega-architecture} and Vega-Lite~\cite{2017-vega-lite}, respectively.
ECharts~\cite{li2018echarts} is also a general visualization grammar.
More grammars target specific domains and visualization types, such as tree visualizations (GoTree~\cite{GoTree}), 
space-filling layouts (HiVE~\cite{Baudel2012, Slingsby2009}), 
maps (Florence~\cite{Poorthuis2020}), 
animations (Canis~\cite{Canis}, Gemini~\cite{Gemini}), 
large datasets (Kyrix~\cite{Kyrix}, Kyrix-S~\cite{Kyrix-S}), tables of counts, proportions, and probabilities (Product Plots~\cite{product_plots}), 
genomics data (Gosling~\cite{Gosling}), and 
interactive data comics~\cite{wang2021interactive}. 
Similar to these grammars, our main contribution is the identification of logic and functional \textit{primitives} alongside ways to combine these primitives. However, none of these grammars explicitly supports networks by lacking graph layouts, matrix orderings, network metrics, graph transformations, clustering, or the ability to construct networks from tabular data.

\textbf{Grammars for network visualization} include ggnetwork~\cite{ggnetwork} and ggraph~\cite{ggraph}, which are extensions to ggplot2 providing a declarative interface for network visualization.
Ggraph provides a function that performs a layout algorithm to assign positions to nodes, and extends ggplot2 by providing additional \texttt{geoms} (geometric objects - i.e., marks) for drawing edges (\texttt{geom\_edges}), nodes (\texttt{geom\_nodes}), and labels (\texttt{geom\_nodetext}, \texttt{geom\_nodelabel}, \texttt{geom\_edgetext} and \texttt{geom\_edgelabel}), and transformations to repel labels to avoid overlaps.
Building on ggplot2, it provides advantages such as allowing the visual properties of nodes and links to be specified in the same way as for other marks in ggplot2 (e.g., the size and color of nodes are specified in the same way as for points in a scatterplot), and inheriting support for faceting.
Ggraph is similar, in that it provides layout functions, and geoms for drawing nodes and edges. 
However, it is broader in scope and besides BioFabric layouts includes \textit{``spatial node layouts such as treemaps, partitions, [...], and circle packing''}~\cite{ggraph_nodes}. 

Both ggraph and ggnetwork provide support for an equivalent to the topological layouts and some of the visual marks in NetPanorama. Yet, they don't directly support network construction and data or network transformations, requiring the use of different R packages. A more fundamental limitation in their expressiveness is that they do not support the representation of nodes or edges by complex glyphs (\autoref{sec:visualencodings}), and have no equivalent of the nested vis blocks in \grammar{} (\autoref{sec:glyphs-and-nesting}). 
A second fundamental limitation is that users write R scripts that rely on a series of function calls (combined using an overloaded \texttt{+} operator) and so they are \textit{embedded} Domain Specific Languages. This embedded nature makes them inherently unsuitable for use as a common representation for visualizations to be used across tools in the way that Vega specifications can be generated using the Python library Altair~\cite{altair}, the R library vegawidget~\cite{vegawidget}, or systems such as Draco~\cite{Draco} or Ivy~\cite{Ivy}.

SetCoLa~\cite{setcola} is a library to express constraints for node-link layouts, such as grouping nodes into sets based on their attributes, aligning nodes horizontally or vertically within a set, position these nodes to the left/right/above/below another set, impose a horizontal or vertical order of node positions, arrange nodes in a circle, try to pull the nodes in a set together into a tight cluster, or exclude non-members of a set from entering the convex hull surrounding the nodes in a set. These constraints allow a range of layouts, which are then calculated by WebCola~\cite{CoLa, WebCoLa}. However, SetCoLa does not support modifying the structure of a network nor specifying aspects of its visual appearance other than the positioning of nodes (in node-link-like layouts). It also does not support matrices and visualizations other than node-link diagrams. \grammar{} seamlessly integrates with SetCoLa.

\grammar{} is best compared to GLO-STIX~\cite{glo-stix} which defines graph visualizations as an ordered set of \textit{graph-level operations} (GLO) for \textit{Specifying Techniques and Interactive eXploration} (STIX) such as `Size Nodes by Count' or `Aggregate Nodes by attribute0 and attribute1'. \grammar{} extends and complements these GLOs (notably graph projection, graph analysis, node and link styling, and layouts) and offers a declarative, rather a procedural way that requires specific operations to happen before others. This allows \grammar{} to express a wider range of designs and techniques. A direct comparison with examples from both languages is found on our website.\footnote{http://netpanorama.netlify.app/docs/example-sequences/les-mis/intro}
\section{Design Criteria}
\label{sec:criteria}

In designing \grammar{}, we set out to create a domain-specific language that can:

\textbf{C1: Describe networks, including multivariate, geographical, and temporal networks}.
Relational data are fundamentally different from tabular data, and many operations for network visualizations are defined in terms of nodes and connections.
A grammar for network visualization must be able to interpret data as networks consisting of nodes linked by links, rather than as just tables. It needs to support links with direction, and with node and link attributes that can be numeric (such as edge weights), categorical (such as link types), temporal (such as times corresponding to each link), or geographic (such as node positions).
While existing grammars and toolkits cover general node and link attributes, geographic and temporal information is currently not supported well (Table \ref{tab:comparison}).

\textbf{C2: Specify a range of visual representations}. 
\grammar{} is designed to be able to create many different 
types of network visualization
for networks; rather than representing each as a fixed template, it represents them as combinations of primitives that can be flexibly combined to create new visualizations, and which allows small modifications to be easily made. 
The choice of functional primitives was informed by considering both the network visualizations that we had worked with ourselves, and those described in the wider literature (such as the reviews~\cite{lee2006task,ahn2013task,STAR_dynamic,geospatial_networks,STAR_multilayer,STAR_multivariate,STAR_large,STAR_groups}).
None of the existing grammars are currently able to create visualizations other than (force-directed other other optimized) node-link diagrams and 
derivatives; our focus is on extending this gap and allowing designers and developers to leverage this huge design space and mix and match designs (\autoref{fig:teaser}).

\textbf{C3: Construct, wrangle and analyze networks}. It is often necessary to define a network from one or more data tables, to calculate network metrics such as node degrees, or to modify a network by projecting, filtering, or aggregating.
Currently, the modeling and analysis are separated from the visualization, with different tools (whether GUI tools, toolkits, or libraries) used for both.
\grammar{} aims to support the \textit{entire} process of network visualization, starting with loading data and constructing a network. This can speed up exploration by avoiding the need to switch between different tools, and allows for network visualizations 
that require modifying a network before visualizing it (such as PivotGraphs~\cite{wattenberg2006visual}, which aggregates nodes based on their attributes).
It also means that the process for generating a visualization is defined by a single file, which can more easily be transferred, annotated, versioned, and inspected for transparency.

\textbf{C4: Integrate with existing (and future)  libraries and toolkits}.
Rather than re-implementing existing functionality, \grammar{} can wrap existing libraries such as WebCola~\cite{WebCoLa} and Tulip~\cite{tulip5} (for layouts), reorder.js~\cite{reorderJS} (for matrix seriation), and some of the libraries that are part of the Vega and Vega-light~\cite{2017-vega-lite} projects (for generic visual marks, scales, and data transformations).
Given the number of network visualization techniques and the dynamic landscape of toolkits, we hope such a unified umbrella approach can provide different functionality to developers and promote more specialized toolkits.

\textbf{C5: Allow interactive exploration and control over styling}. 
\grammar{} aims to provide simple mechanisms for the viewers of visualizations to alter graphical choices and visual mappings (such as the colors, opacity, or size of marks) using controls such as sliders. We also support interactive exploration through highlighting nodes and links, or displaying information with tooltips, and highlighting neighbors or the shortest path between two nodes (which is an inherently network-specific interaction).

\def\checkmark{\textcolor{white}{\tikz\fill[scale=0.2](0,.35) -- (.25,0) -- (1,.7) -- (.25,.15) -- cycle;}}
\def\rot{\rotatebox}

\begin{table}[]
\footnotesize
\centering
\scalebox{0.9}{
\begin{tabular}{p{.1cm}p{2.2cm}|p{.1cm}p{.1cm}|p{.1cm}p{.1cm}p{.1cm}|p{.1cm}p{.1cm}p{.1cm}p{.1cm}|p{.1cm}p{.1cm}}
 &
   &
   \small{\textbf{General}} &
   &
    \small{\textbf{Wrangling}} &
   &
   &
    \multicolumn{4}{l}{\small{\textbf{Libraries}}} &
    \multicolumn{2}{l}{\small{\textbf{DSLs}}} \\
 &
   &
  \rot{90}{\textbf{Netpanorama}} &
  \rot{90}{Vega-Lite} &
  \rot{90}{Ploceus} &
  \rot{90}{Origraph} &
  \rot{90}{Orion} &
  \rot{90}{cystoscape.js} &
  \rot{90}{sigma.js} &
  \rot{90}{G6} &
  \rot{90}{ggnetwork} &
  \rot{90}{GLO-STIX} &
  \rot{90}{GraphVis} \\
  \hline
   \hline
 C1: &
   Network data &
  \cellcolor[HTML]{888888}\checkmark &
  \cellcolor[HTML]{ffffff}- &
  \cellcolor[HTML]{aaaaaa}\checkmark &
  \cellcolor[HTML]{aaaaaa}\checkmark &
  \cellcolor[HTML]{aaaaaa}\checkmark &
  \cellcolor[HTML]{aaaaaa}\checkmark &
  \cellcolor[HTML]{aaaaaa}\checkmark &
  \cellcolor[HTML]{aaaaaa}\checkmark &
  \cellcolor[HTML]{aaaaaa}\checkmark &
  \cellcolor[HTML]{aaaaaa}\checkmark &
  \cellcolor[HTML]{aaaaaa}\checkmark \\
 &
  Time semantics &
  \cellcolor[HTML]{888888}\checkmark &
  \cellcolor[HTML]{aaaaaa}\checkmark &
  \cellcolor[HTML]{ffffff}- &
  \cellcolor[HTML]{ffffff}- &
  \cellcolor[HTML]{aaaaaa}\checkmark &
  \cellcolor[HTML]{ffffff}- &
  \cellcolor[HTML]{ffffff}- &
  \cellcolor[HTML]{ffffff}- &
  \cellcolor[HTML]{ffffff}- &
  \cellcolor[HTML]{ffffff}- &
  \cellcolor[HTML]{ffffff}- \\
 &
  Maps&
  \cellcolor[HTML]{888888}\checkmark &
  \cellcolor[HTML]{aaaaaa}\checkmark &
  \cellcolor[HTML]{ffffff}- &
  \cellcolor[HTML]{ffffff}- &
  \cellcolor[HTML]{ffffff}- &
  \cellcolor[HTML]{ffffff}- &
  \cellcolor[HTML]{ffffff}- &
  \cellcolor[HTML]{ffffff}- &
  \cellcolor[HTML]{ffffff}- &
  \cellcolor[HTML]{ffffff}- &
  \cellcolor[HTML]{ffffff}- \\
 &
  Geocoding &
  \cellcolor[HTML]{888888}\checkmark &
  \cellcolor[HTML]{ffffff}- &
  \cellcolor[HTML]{ffffff}- &
  \cellcolor[HTML]{ffffff}- &
  \cellcolor[HTML]{ffffff}- &
  \cellcolor[HTML]{ffffff}- &
  \cellcolor[HTML]{ffffff}- &
  \cellcolor[HTML]{ffffff}- &
  \cellcolor[HTML]{ffffff}- &
  \cellcolor[HTML]{ffffff}- &
  \cellcolor[HTML]{ffffff}- \\
  &
  Node and link types \& parallel links &
  \cellcolor[HTML]{888888}\checkmark &
  \cellcolor[HTML]{ffffff}- &
  \cellcolor[HTML]{aaaaaa}\checkmark &
  \cellcolor[HTML]{aaaaaa}\checkmark &
  \cellcolor[HTML]{aaaaaa}\checkmark &
  \cellcolor[HTML]{aaaaaa}\checkmark &
  \cellcolor[HTML]{aaaaaa}\checkmark &
  \cellcolor[HTML]{aaaaaa}\checkmark &
  \cellcolor[HTML]{aaaaaa}\checkmark &
  \cellcolor[HTML]{aaaaaa}\checkmark &
  \cellcolor[HTML]{aaaaaa}\checkmark \\
\hline
C2: &
  Other than Node-link Diagrams &
  \cellcolor[HTML]{888888}\checkmark &
  \cellcolor[HTML]{ffffff}- &
  \cellcolor[HTML]{aaaaaa}\checkmark &
  \cellcolor[HTML]{ffffff}- &
  \cellcolor[HTML]{aaaaaa}\checkmark &
  \cellcolor[HTML]{ffffff}- &
  \cellcolor[HTML]{ffffff}- &
  \cellcolor[HTML]{aaaaaa}\checkmark &
  \cellcolor[HTML]{ffffff}- &
  \cellcolor[HTML]{aaaaaa}\checkmark &
  \cellcolor[HTML]{ffffff}- \\
 &
  Node \& link glyphs &
  \cellcolor[HTML]{888888}\checkmark &
  \cellcolor[HTML]{ffffff}- &
  \cellcolor[HTML]{ffffff}- &
  \cellcolor[HTML]{ffffff}- &
  \cellcolor[HTML]{ffffff}- &
  \cellcolor[HTML]{aaaaaa}\checkmark &
  \cellcolor[HTML]{ffffff}- &
  \cellcolor[HTML]{ffffff}- &
  \cellcolor[HTML]{ffffff}- &
  \cellcolor[HTML]{ffffff}- &
  \cellcolor[HTML]{ffffff}- \\
 &
  Facetting &
  \cellcolor[HTML]{888888}\checkmark &
  \cellcolor[HTML]{aaaaaa}\checkmark &
  \cellcolor[HTML]{aaaaaa}\checkmark &
  \cellcolor[HTML]{ffffff}- &
  \cellcolor[HTML]{aaaaaa}\checkmark &
  \cellcolor[HTML]{ffffff}- &
  \cellcolor[HTML]{ffffff}- &
  \cellcolor[HTML]{ffffff}- &
  \cellcolor[HTML]{aaaaaa}\checkmark &
  \cellcolor[HTML]{ffffff}- &
  \cellcolor[HTML]{ffffff}- \\
  \hline
C3: &
  Multiple tables &
  \cellcolor[HTML]{888888}\checkmark &
  \cellcolor[HTML]{aaaaaa}\checkmark &
  \cellcolor[HTML]{ffffff}- &
  \cellcolor[HTML]{aaaaaa}\checkmark &
  \cellcolor[HTML]{aaaaaa}\checkmark &
  \cellcolor[HTML]{ffffff}- &
  \cellcolor[HTML]{ffffff}- &
  \cellcolor[HTML]{ffffff}- &
  \cellcolor[HTML]{ffffff}- &
  \cellcolor[HTML]{ffffff}- &
  \cellcolor[HTML]{ffffff}- \\
 &
  Network transform. &
  \cellcolor[HTML]{888888}\checkmark &
  \cellcolor[HTML]{ffffff}- &
  \cellcolor[HTML]{aaaaaa}\checkmark &
  \cellcolor[HTML]{aaaaaa}\checkmark &
  \cellcolor[HTML]{aaaaaa}\checkmark &
  \cellcolor[HTML]{aaaaaa}\checkmark &
  \cellcolor[HTML]{ffffff}- &
  \cellcolor[HTML]{aaaaaa}\checkmark &
  \cellcolor[HTML]{ffffff}- &
  \cellcolor[HTML]{aaaaaa}\checkmark &
  \cellcolor[HTML]{ffffff}- \\
 &
  Network metrics &
  \cellcolor[HTML]{888888}\checkmark &
  \cellcolor[HTML]{ffffff}- &
  \cellcolor[HTML]{aaaaaa}\checkmark &
  \cellcolor[HTML]{ffffff}- &
  \cellcolor[HTML]{aaaaaa}\checkmark &
  \cellcolor[HTML]{aaaaaa}\checkmark &
  \cellcolor[HTML]{ffffff}- &
  \cellcolor[HTML]{ffffff}- &
  \cellcolor[HTML]{ffffff}- &
  \cellcolor[HTML]{ffffff}- &
  \cellcolor[HTML]{ffffff}- \\
&
  Node seriation &
  \cellcolor[HTML]{888888}\checkmark &
  \cellcolor[HTML]{ffffff}- &
  \cellcolor[HTML]{ffffff}- &
  \cellcolor[HTML]{ffffff}- &
  \cellcolor[HTML]{aaaaaa}\checkmark &
  \cellcolor[HTML]{ffffff}- &
  \cellcolor[HTML]{ffffff}- &
  \cellcolor[HTML]{ffffff}- &
  \cellcolor[HTML]{ffffff}- &
  \cellcolor[HTML]{ffffff}- &
  \cellcolor[HTML]{aaaaaa}\checkmark \\
  \hline
C5: &
  Interactive exploration. &
  \cellcolor[HTML]{888888}\checkmark &
  \cellcolor[HTML]{aaaaaa}\checkmark &
  \cellcolor[HTML]{aaaaaa}\checkmark &
  \cellcolor[HTML]{aaaaaa}\checkmark &
  \cellcolor[HTML]{aaaaaa}\checkmark &
  \cellcolor[HTML]{aaaaaa}\checkmark &
  \cellcolor[HTML]{aaaaaa}\checkmark &
  \cellcolor[HTML]{aaaaaa}\checkmark &
  \cellcolor[HTML]{ffffff}- &
  \cellcolor[HTML]{ffffff}- &
  \cellcolor[HTML]{ffffff}- \\
 & Interactive styling &
  \cellcolor[HTML]{888888}\checkmark &
  \cellcolor[HTML]{aaaaaa}\checkmark &
  \cellcolor[HTML]{ffffff}- &
  \cellcolor[HTML]{ffffff}- &
  \cellcolor[HTML]{ffffff}- &
  \cellcolor[HTML]{ffffff}- &
  \cellcolor[HTML]{ffffff}- &
  \cellcolor[HTML]{ffffff}- &
  \cellcolor[HTML]{ffffff}- &
  \cellcolor[HTML]{ffffff}- &
  \cellcolor[HTML]{ffffff}- \\
\hline
\end{tabular}%
}
\caption{Comparing \grammar{} to other toolkits and grammars.
}
\vspace{-1.5em}
\label{tab:comparison}
\end{table}

\section{\grammar}
\label{sec:concepts}

\begin{figure}
    \centering
    \includegraphics[width=1\columnwidth]{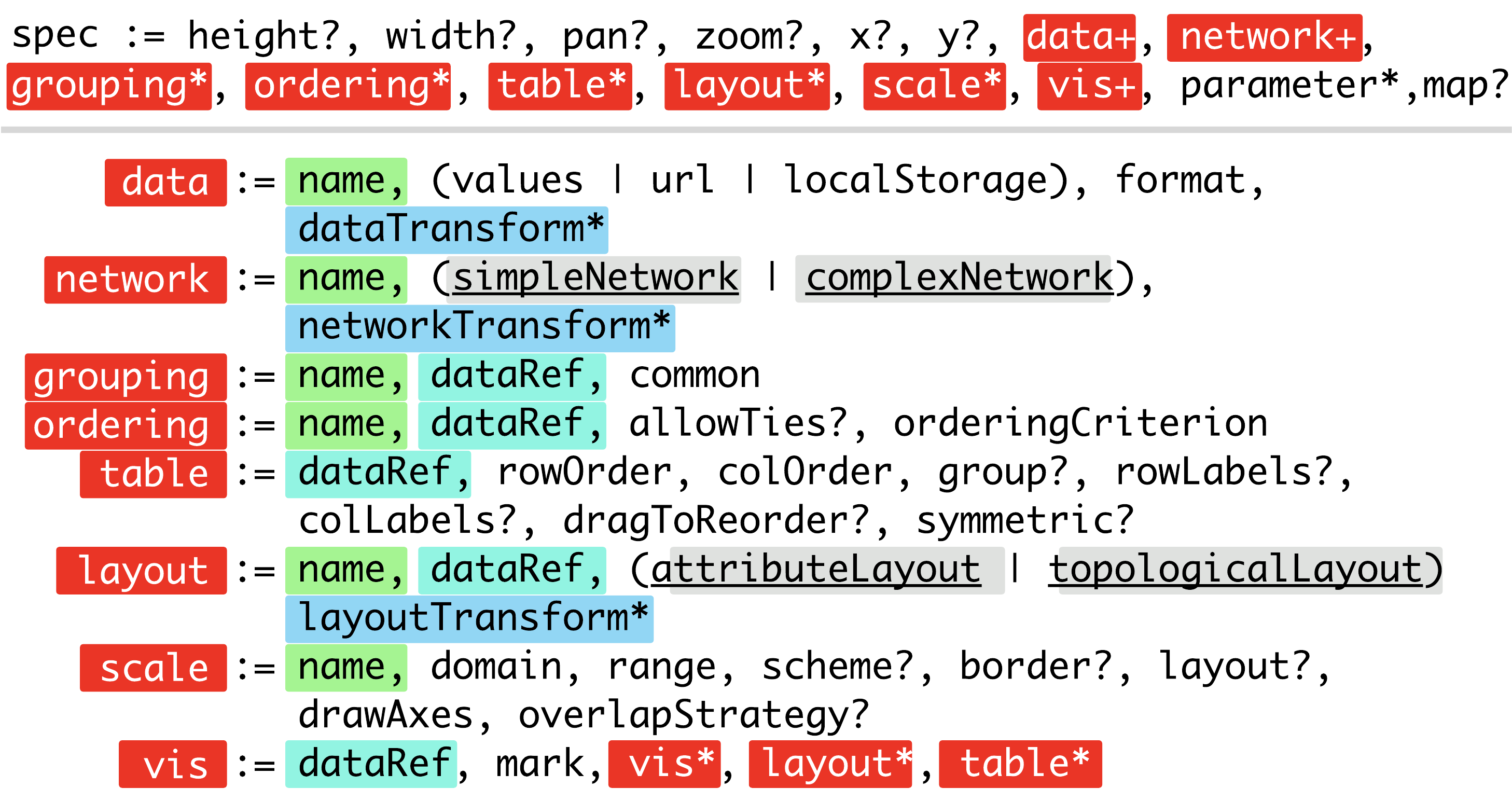}
    \caption{
    Top-level constructs in \grammar{} (red); apart from \codeown{vis}, instances of these have a \green{name} and may point to the name of another object (dataRef) such as a network or a data array.
    Instances of \codeown{data}, \codeown{network}, and \codeown{layout} can be modified by \blue{transform}s.
    As well as being defined at the top-level of a specification, instances of a \codeown{vis}, \codeown{layout}, and \codeown{table} can be \textit{nested} within the definition of a \codeown{vis} instance.
    Gray text indicates alternative forms of a construct.
    }
    \vspace{-2em}
    \label{fig:spec}
\end{figure}

\grammar{} specifies a network visualization as a combination of \textit{constructs} of different types (\ca{data, network, grouping, ordering, table, layout, scale, vis}; see \autoref{fig:spec}). 
Apart from \codeown{vis}, instances of these have a name that can be referred to elsewhere in the specification; they take a data file or an in-memory object as input and either modify this in-memory object, or create a new in-memory object.
This allows structures along the visualization process, e.g., data tables, layouts, orderings, and different network models, to co-exist in parallel, to be displayed at the same time, or to be extended by different constructs. 
Instances of \codeown{data}, \codeown{network}, and \codeown{layout} can be modified by transform constructs, which are \textit{local}: they do not produce named output, and act to modify a specific object.

In this section, we discuss each top-level construct individually and refer to detailed grammar structure where appropriate. For simplicity, when referring to syntax in the paper, we use \codeown{red} for constructs and attributes, and \codeownv{teal} for attribute values. To express quantity and option, we use: \texttt{*} zero or more, \texttt{+} at least one, \texttt{?} optional, and \texttt{|} alternatives. Note that some of the constructs in \autoref{fig:spec} are actually arrays of arbitrarily many constructs (\texttt{*,+}).
The full documentation of the grammar and all options can be found online along examples: \url{https://netpanorama.netlify.app}. 

\subsection{Data Import and Transformation}

The \codeown{data} construct loads tabular data in a given format (e.g., CSV or JSON) from a location (URL, inline JSON literal, or from browser \texttt{localStorage}). Once loaded, data is transformed into an internal data model and can be manipulated by 
data \codeown{transformation}s: 
filtering, adding and removing columns, normalizing, summing values, binning, aggregating rows, and performing calculations based on expressions. 
Transformation can also ensure consistent identifiers for nodes across different tables; especially if tables come from different 
databases or sources. For example, identifiers may have a prefix such as \texttt{doi:} or \texttt{isbn:} in one location but not another. \grammar{} supports \textbf{geocoding} 
to automatically assign geographic coordinates from place names using the GeoCode API.\footnote{\url{https://geocode.maps.co}}

\subsection{Networks}  

\begin{figure}[h!]
    \centering
\includegraphics[width=1\columnwidth]{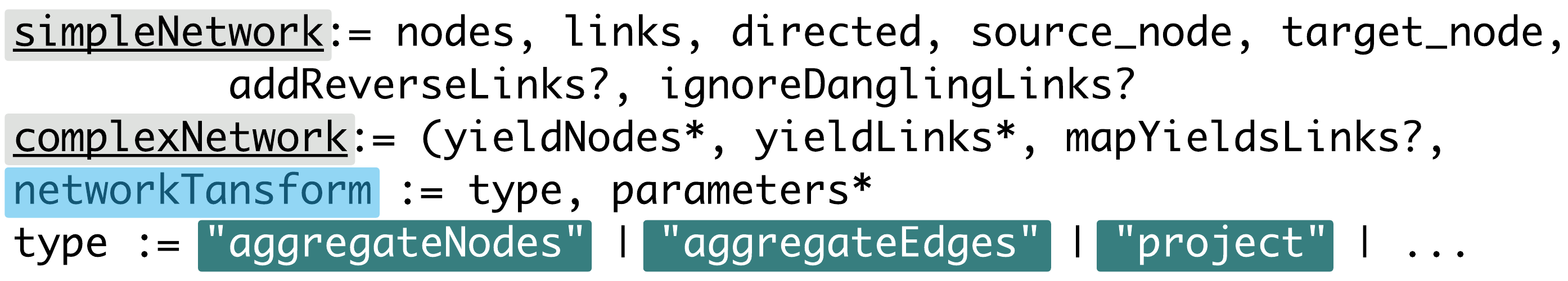}
    \caption{Network construction and topological transformations.
    Different types of transformation are distinguished by their \texttt{type}, and accept different sets of parameters.
    }
    \label{fig:spec-network}
\vspace{-1.8em}
\end{figure}

Once data tables or other formats are loaded, the \codeown{network} construct creates networks by defining the semantics of the data in terms of nodes, links, and their attributes. Since \ca{network} really is an array, \grammar{} can define as many networks as wanted, using the same data or parts of it. The network topology can then be used to calculate metrics and clusters (\autoref{sec:metrics}), layouts (\autoref{sec:layouts}) and graph-specific queries such as finding common neighbors and paths between nodes (\autoref{sec:transformations}). Network construction is trivial in cases where nodes and links are already clearly defined in the input data file format (e.g., for Pajek or GraphML files which can be directly loaded inside a \ca{network} construct rather than \ca{data}).  
In cases when a network is constructed from data tables,  rows and columns in those tables need to be mapped to nodes, links, and attributes in different ways, either 
\textit{i)} using only a subpart of the data in a given network (or tables), 
\textit{ii)} creating multiple networks from the same data, and 
\textit{iii)} creating networks from multiple input files (including tables). 
If only one table is imported, a \texttt{simpleNetworkSpec} maps column names to network attributes (\autoref{fig:spec-network}). If more than one table is used to create a network, 
a \texttt{complesNetworkSpec} defines nodes and links through the \codeown{yieldNodes} and \codeown{yieldLinks} constructs and assigns column semantics through \codeown{source\_id}, \codeown{target\_id} across multiple tables (\autoref{fig:spec-network}).

\subsection{Network Metrics}
\label{sec:metrics}

Metrics are calculated within a \codeown{networkTransform} construct, specified by its type and potentially other parameters (\autoref{fig:spec-network}). Certain visualization and interaction techniques require metrics, e.g., to scale, color, or order (\autoref{sec:orderings}), filter, or cluster nodes.
\grammar{} can calculate metrics (diverse degree measures, in-betweenness centrality, closeness, eccentricity~\cite{centrality_measures}), perform clustering, or aggregating information from a node's neighbors through an \ca{attributeSelector} attribute on the \ca{networkTransform}.
When calculating clusters, a clustering method is chosen (\ca{method}), e.g., \cv{louvain}, alongside a new data field \ca{as} where to store a cluster number for each node. This method allows the computation (and visualization) of multiple clusterings.

\subsection{Topological Network Transformations}
\label{sec:transformations}

Networks imported from existing data sets might require further refinements for a given visualization (e.g., grouping nodes of a specific topic), or for a task (e.g., reducing network size). Topological transformations alter the topology of a network by adding, removing, or replacing nodes and links. Topological network transformations are declared individually for each \codeown{network}, within its \codeown{networkTransform} construct (alongside metrics). 
A transform has a \ca{type} attribute alongside transformation specific 
attributes. Available transforms include:
\begin{itemize}[noitemsep,leftmargin=*]
    \item \textbf{filtering nodes and links} based on calculated measures, data attributes, or a node's  connectivity (\codeowna{type}{("removeIsolated"|"filterNodes"|"filterEdges")}).
    
    \item \textbf{adding nodes and links}, e.g., to create a node for each attribute (field) value found on a node (\codeowna{type}{"promote"}, \codeowna{field}{<string>}) or connect all nodes with a specific type by a new type of link (\codeowna{type}{"connect"}, \codeowna{field}{<string>}, \codeowna{as}{<string>}).
        
    \item \textbf{aggregating nodes and/or links} into a meta node based on similar values for one or more attributes, e.g., cluster number or degree, or any other attribute value: \codeowna{type}{("aggregateNode"| "aggregateEdges")}, \codeowna{type}{<field(s)>}).
    
    \item \textbf{projecting edges} replaces a path between two nodes via a third node with a single link directly joining them (i.e., $A\rightarrow B\rightarrow C$ becomes $A \rightarrow C$). The transformation operates on the entire network and takes two parameters: the type of nodes that should remain, and the type removed by the projection. 
    (\codeowna{type}{"project"}, \codeowna{remainingNodeType}{<string>}, \codeowna{removedNodeType}{<string>}). A dedicated \codeown{linkDataReducer} specifies how to aggregate the data from the links to be merged.
\end{itemize}

Each transformation requires its own set of parameters, many of which can be expressed through \codeown{expressions} similar to Vega-Lite.
Unlike other constructs in \grammar{}, the operations construct is an ordered array which functions as a pipeline of operations with metrics and transformations executed in their exact order. Similar to network definitions in GLO-STIX, this allows taking the results of an operation as input for the next operation. 
An example of this is demonstrated in \autoref{fig:network_transformations}: calculate node degrees, filter nodes with degree $>2$, and calculate a clustering, then replace clusters by meta-nodes and links between clusters by meta-links:

\begin{figure}[h!]
    \includegraphics[trim={0 0 0 16.5cm},clip,width=.48\textwidth]{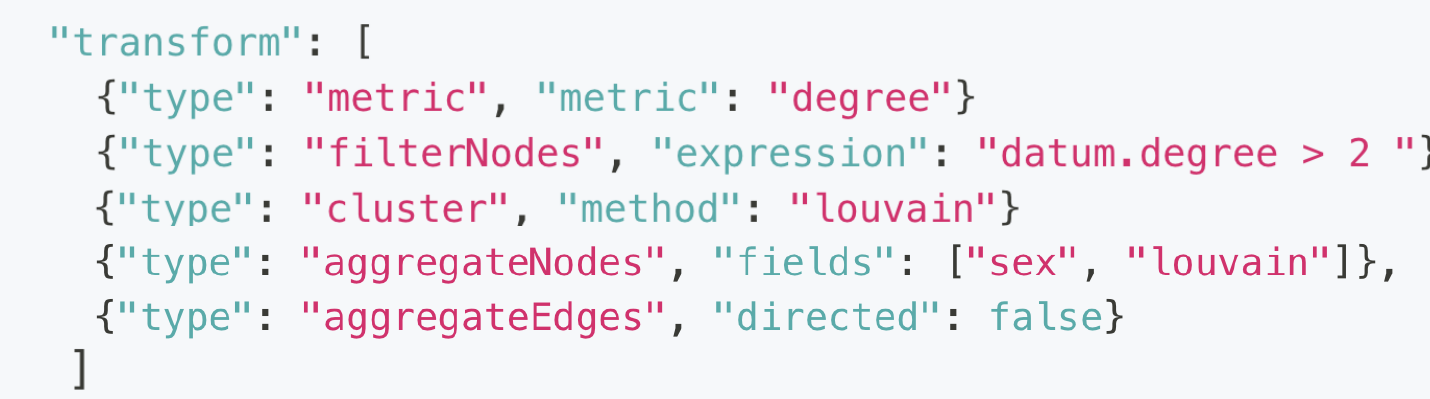}
    \caption{Examples of network transformations specified in \grammar{}. These transforms are defined in a \codeown{transform} attribute of the definition for the corresponding \codeown{network}.  }
    \vspace{-1.8em}
    \label{fig:network_transformations}
\end{figure}

\subsection{Seriations and orderings}
\label{sec:orderings}

Many types of layout depend on an ordering over the elements that they position (e.g., adjacency matrices, arc diagrams, hive-plots, circular layouts, parallel edge-splatting~\cite{burch2011parallel} or jigsaws~\cite{stasko2008jigsaw}; see \autoref{sec:layouts}).
In \grammar, orderings for nodes, groups, and other data elements are defined in the top-level \codeown{orderings} construct and can be accessed by the visual marks for placement (\autoref{sec:visualencodings}). 
\textbf{Attribute orderings} order nodes based on a data attribute value or a metric (specified as \ca{orderingCriterionField} which can result from a transformation such as degree. A \textbf{matrix seriation} optimizes node order based on their connectivity in order to reveal patterns in an adjacency matrix~\cite{behrisch2016matrix}. Seriation requires a method and a distance (\codeowna{method}{("optimal-leaf-order"|
"barycentre"|"bandwidth-reduction"|"pca")}) and a \codeowna{dis\-tance}{"euclidean"|"manhattan"|"minkowski"|"chebyshev"
|"hamming"|"jaccard"|"braycurtis"}).

While originating in the context of adjacency matrices, \grammar{} makes it possible to use the same seriation across all visualization designs that offer some sort of linear ordering for nodes: arc diagrams, adjacency lists, hive plots, jig-saw, etc. In our implementation, \grammar{} stores an ordering as a mapping from object \texttt{id}s to a number to allow for co-existence of several orderings on the same data set and enable, e.g., interactive switching between orderings (see \autoref{sec:interaction}) to explore patterns in matrices and other visualizations~\cite{behrisch2016magnostics}.

\subsection{Layouts}
\label{sec:layouts}

\begin{figure}[h!]
    \centering
    \includegraphics[width=1\columnwidth]{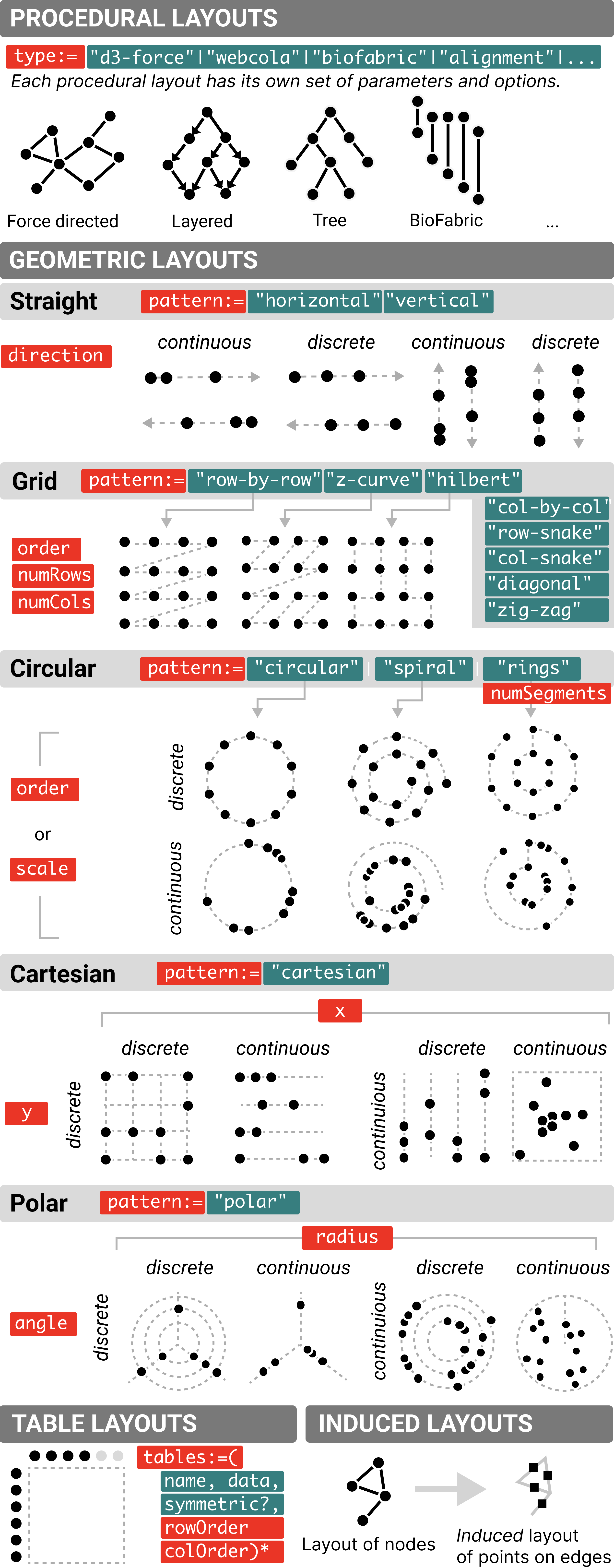}
    \caption{Layout overview. Whether a dimension of a layout is discrete or continuous depends on what the \ca{scale} or \ca{order} that is used: an \ca{order} will always lead to a discrete version, a \ca{scale} can lead to either.}
    \label{fig:layouts}
    \vspace{-1.5em}
\end{figure}

In the absence of a comprehensive taxonomy of general layout techniques, the challenge for a network visualization grammar is to express the many diverse layouts in a coherent way, including other than the classical node-link layouts (e.g.,~\cite{Tamassia2016-la}). 
We categorized layouts
into \textit{geometric layouts}, \textit{procedural layouts}, \textit{table layouts}, and \textit{induced} layouts (see the overview in \autoref{fig:layouts}). Most of these layouts are defined inside the \codeown{layout} array construct, 
with the exception of tables (which combine a layout and a grouping, and are defined in a separate \ca{table} construct - \autoref{sec:table-layouts}) and induced layouts (which are defined alongside the visual mark that they depend on - \autoref{sec:induced-layouts}).

\subsubsection{Procedural layouts}
Procedural layouts position nodes based on the network topology and calculations on the entire network structure (which are often iterative). 
Common examples of procedural layouts include force-directed layouts, layered layouts for directed acyclic graphs, and tree layouts. \grammar{} refers to procedural layout implementation by their name, while allowing for parameterizing each specific layout: \codeowna{type}{("d3-force"|"webcola"|"tulip"|"cytoscape"|...)} with a respective set of individual options (e.g., \ca{forces}). 

\subsubsection{Geometric layouts}
Geometric layouts arrange nodes or groups according to geometric structures and rules that can be independent of the network topology. 
They position elements along their respective geometric structures by using either an \ca{order} or a \ca{scale} applied to their attributes (data attributes, calculated metrics, or geographic locations).
Whether a dimension of a layout is continuous or discrete depends on the scale or order that is provided: if an \ca{order} or discrete \ca{scale} is used, then the elements will be positions at discrete steps; if a continuous \ca{scale} is used, then they will be position continuously.
To \textbf{avoid node overlap} in situations where multiple nodes end up at the same positions, \grammar{} provides \ca{layoutTransforms} that apply overlap removal techniques (\codeowna{type}{"offset"|"jitter"|"beeswarm"}).

\grammar{} distinguishes between the following types of geometric layouts (\autoref{fig:layouts}), specified in the \ca{pattern} attribute of a \ca{layout} construct:
\begin{itemize}[noitemsep,leftmargin=*]

    \item \textbf{Straight geometric layouts:} nodes or links are laid out on a single straight line  (\codeowna{pattern}{("horizontal"|"vertical")}).
    A  common use of line layouts to draw an Arc Diagram~\cite{wattenberg2002arc}.

    \item \textbf{Grid layouts:} nodes and links are laid out in a linear order along a curve filling a grid, based on an \ca{ordering}. Grid layouts can be used to create concisely represent large networks or
    to lay out nodes inside clusters. \grammar{} provides different strategies for building a grid, including \cv{z-curves}, \cv{Hilbert} curves, and \cv{zig-zag} curves. Besides large networks, another use case of grid layouts can be to assess correlation between a node attribute and either another attribute ordering or a matrix seriation (topological similarity); this functionality is implemented in Tulip~\cite{tulip5}. 
    
    \item \textbf{Circular geometric layouts:} nodes or links are laid out in a circle based on an \ca{order} or \ca{scale}: either a single circle (\cv{circular}), a \cv{spiral}, or in \cv{rings} in which a continuous order jumps to the next level of nested circle when a circle is filled. 
    Circular layouts are popular in neuroscience and can prevent visual clutter while showing all nodes in a equal manner. Rings can separate nodes with different attributes

    \item \textbf{Cartesian geometric layouts:} nodes are laid out in a 2-dimensional orthogonal space like in a scatterplot, according to two node attributes (\codeowna{x}{\{field:<string>\}} and \codeowna{y}{\{field:<string>\}}). For example, a layout using two discrete attributes is used to render a PivotGraph~\cite{wattenberg2006visual}. A version with two continuous attributes is used to render scatterplots as in GraphDice~\cite{bezerianos2010graphdice} or position nodes based on their geographic coordinates.
    
    \item \textbf{Polar geometric layouts:} similar to Cartesian layouts, nodes or links are laid out according to two dimensions, but
    one coordinate determines \ca{angle} and the other radial \ca{distance} from the central point. 
    An example use for a polar layout with a continuous radius but a discrete angle is rendering Hive plots~\cite{krzywinski2012hive}.

\end{itemize}

\subsubsection{Table layouts}\label{sec:table-layouts} 
Table layouts position elements along rows and columns of a table, based on the value of two specified fields or expressions (\codeown{rowOrder} and \codeown{colOrder}); this also implicitly defines a grouping, with elements grouped together if they have the same value for both fields. 
Unlike the other layout types, a table is thus a construct that simultaneously defines a grouping and a layout: when used to define a visual mapping, it both defines the data to be mapped to marks (the groups corresponding cells in the table) and their positions.
A typical use of a table layout is to create an adjacency matrix, in which edges are positioned into rows and columns based on their source and target nodes, and edges with the same source and target are grouped together in the same cell.
Similar to geometric layouts, row and column orders can be defined based on attributes (\codeown{order}) or network topology (seriation~\cite{behrisch2016matrix}).

\subsubsection{Induced layouts}\label{sec:induced-layouts}
Induced layouts are layouts that are derived from the positions at which other visual marks are drawn.
For example, we could define an induced layout that records the positions of the mid-point of the links in a node-link diagram as they are drawn, and then use this induced layout to position glyphs that show information about the attributes of each link.
In this example, the induced layouts depend not only on another \codeown{layout} (the layout of the nodes), but also on the properties of the \codeown{linkPath} mark (such as its shape and degree of curvature); hence the induced layout is defined within the \codeown{vis} construct (Section \ref{sec:visualencodings}) for the links, rather than alongside the other layouts in the \codeown{layout} construct.

\subsection{Visual Encodings}
\label{sec:visualencodings}

The \codeown{vis} construct defines mappings from data elements (nodes, links, groups) referenced in a \ca{dataRef} attribute and their data attributes 
to visual \ca{marks} and their visual attributes.
A \ca{mark} in \grammar{} is similar to a \texttt{mark} in Vega-Lite: 
mark types
include lines, bars, areas underneath lines, text, and symbols (including squares and circles). Visual properties of these marks can be specified as a constant value, a data attribute (typically transformed by a \codeown{scale} function or an expression), or a parameter bound to an input control (Section \ref{sec:interaction}).
However, \grammar{} also includes an additional \codeown{linkpaths} visual mark to render links between two nodes.
A \codeown{vis} construct can include an arbitrary number of marks to represent the same data element, facilitating the use of glyphs to render more complex information about nodes, links, and groups of either.

\subsubsection{LinkPaths}

Positions for each visual element are internally read from the \codeown{layout} parameter that references a \ca{layout} or a \ca{table}. For linkpaths representing links, start and end coordinates are taken from the respective incident nodes. \codeown{Linkpath}s can be rendered as one of the following shapes which we found across the literature: \codeowna{type}{("line"|"arc"|"curve")}. Vega's linkpath version is much more limited version of this as a linkpath \textit{transform}; we consider it appropriate in the context of networks to handle link shapes as a fundamental mark type, rather than as a transform. 

\codeown{Linkpath} directionality is visually indicated by line-end markers such as filled arrow-heads, color gradients, changes in line width to create a wedge shape)~\cite{Holten_2006}, asymmetric curvature~\cite{fekete:hal-00875194}, half-lines that start at one node but do not extend the full distance to the other node~\cite{Becker_1995}, or an asymmetric envelope applied to the sinusoidal patterns as in the ABySS-Explorer~\cite{abyss_explorer}. An alternative to geometric \codeown{linkpath}s is a series of \codeowna{shape}{("wedge"|"line"|"glyphs")} along the path that a curve would take 
and by specifying a \codeown{glyphShape} alongside a \codeown{separation}. 
\grammar{} also supports strictly parallel or slightly curved (bundled) lines for \codeown{linkpath} links between the same pairs of nodes with the \codeown{parallelLinks} parameter specified with a \codeowna{type}{("line"|"curve")} and a \codeowna{gap}{<number>}  optional parameter.

\begin{figure}[h!]
    \centering
    \vspace{-1em}
    \includegraphics[width=0.8\columnwidth]{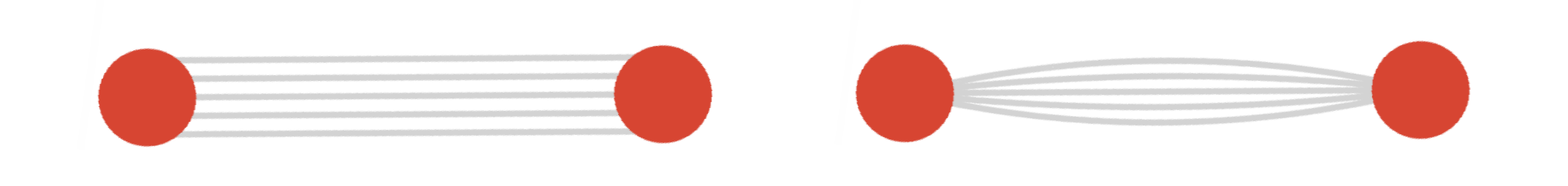}    
    \caption{Different link renderings and gaps.}
    \vspace{-1.5em}
\end{figure}

\subsubsection{Glyphs}
\label{sec:glyphs-and-nesting}

\begin{figure}[h]
    \centering
    \vspace{-1em}
    \includegraphics[width=1\columnwidth]{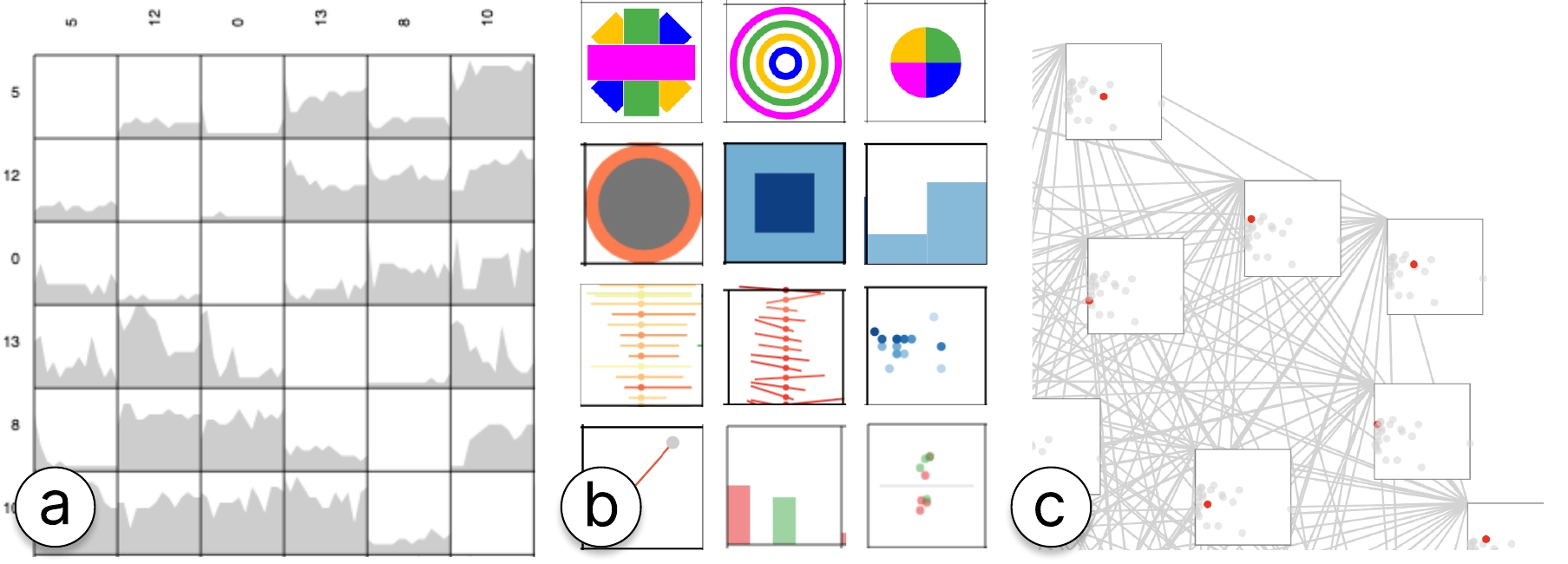}
    \caption{Glyphs and shapes:
    (a) an adjacency matrix containing area charts showing link data,   
    (b) 12 different glyph designs that could be used in an adjacency matrix, 
    (c) scatterplot glyphs showing multivariate node data on a node-link diagram.
    }
    \label{fig:glyphs-example}
    \vspace{-1em}
\end{figure}

\grammar{} also provides ways to create more complex visual marks composed of several other visual marks. Such glyphs have been frequently used to show multiple data attributes on nodes and links such as time~\cite{brandes2011asymmetric} but can also be used to show information about nodes in a cluster~\cite{henry2007nodetrix,ontotrix}. Glyphs can be created using several mechanisms:

\begin{itemize}[noitemsep,leftmargin=*]
\item \textbf{\codeown{area} marks} can create arbitrary shapes as a single mark. This allows a line-graph to be created as a single mark (e.g., to show a time-series, as in \autoref{fig:glyphs-example}(a)).

\item 
A \codeown{vis} block can define \textbf{multiple marks} to be created from the same entries and layouts: by applying appropriate offsets, this allows creation of bar charts, overlaid shapes or complex stroke-figures as introduced by Brandes and Nick~\cite{brandes2011asymmetric} (Figure \ref{fig:glyphs-example}(b)).

\item 
\textbf{nesting} multiple \codeown{vis} constructs with their own \ca{entries} can specify a visual encoding to be applied to each individual element in each of the different entries. This can be used to create \textbf{small multiples} (Figure \ref{fig:nesting}) or glyphs within the cells of a  matrix (Figure \ref{fig:glyphs-example}(a-b)). 
\end{itemize}

\subsubsection{Facetting}

\begin{figure}[]
    \centering
\includegraphics[width=1\columnwidth]{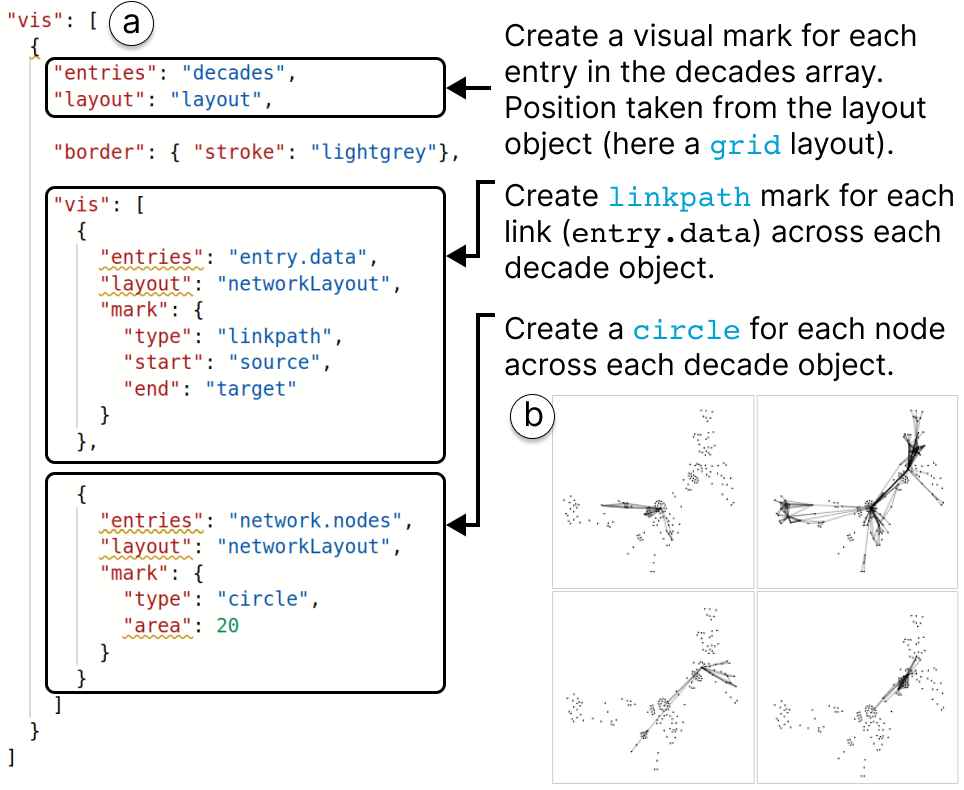}
    \caption{Examples of a nested visualization that draws small multiples of a node-link diagram, one for each decade in the data set. The example requires links to be grouped by decade first (producing a grouping called \texttt{decades}). Node-link diagrams are laid out in using a \textit{grid} layout. 
    } 
    \label{fig:nesting}
    \vspace{-1.5em}
\end{figure}

Small multiples are commonly used to visualize networks over time~\cite{beck2017taxonomy} or facet by other node and link attributes~\cite{ploceus}. To create a small multiple for each year in a dynamic network (Figure \ref{fig:nesting}), we first group nodes and links by their year, 
then create one \codeown{vis} for each of these groups and eventually create separate nested \codeown{vis} for nodes and links. Within the inner \codeown{vis}, it is possible to determine the available space of the small multiple as \codeown{bounds} and scale the content appropriately. Both of the inner \ca{vis} constructs (which render the links and nodes) use the same (nodelink) layout, while the root \ca{vis} references a grid layout to place individual multiples.

Vega-Lite provides \texttt{rows} and \texttt{col} attributes as a shortcut to creating small multiples, however while semantically similar, our version is more expressive since it allows arbitrary layouts of small multiples as well as creating small multiples entirely free from attributes, such as different types of layouts for the same network, different networks side by side, or different visual designs juxtaposed.

\subsection{Node label placement and visibility}
\label{sec:labeling}

Label placement is a challenge for visualization in general, but comes with additional challenges in network visualization because most of the screen-space is usually used to show nodes, links, maps, or time.

\grammar{} offers  \textit{generic} features for 
\textit{a)} 
fixing certain label's position when panning and zooming (explained in the context of interaction in Section \ref{sec:interaction}), and 
\textit{b)} 
showing labels based on node attributes or metrics. All those options are specified inside the respective \ca{mark} representing the label text.

\begin{figure}[t]
    \centering
    \includegraphics[width=1\columnwidth]{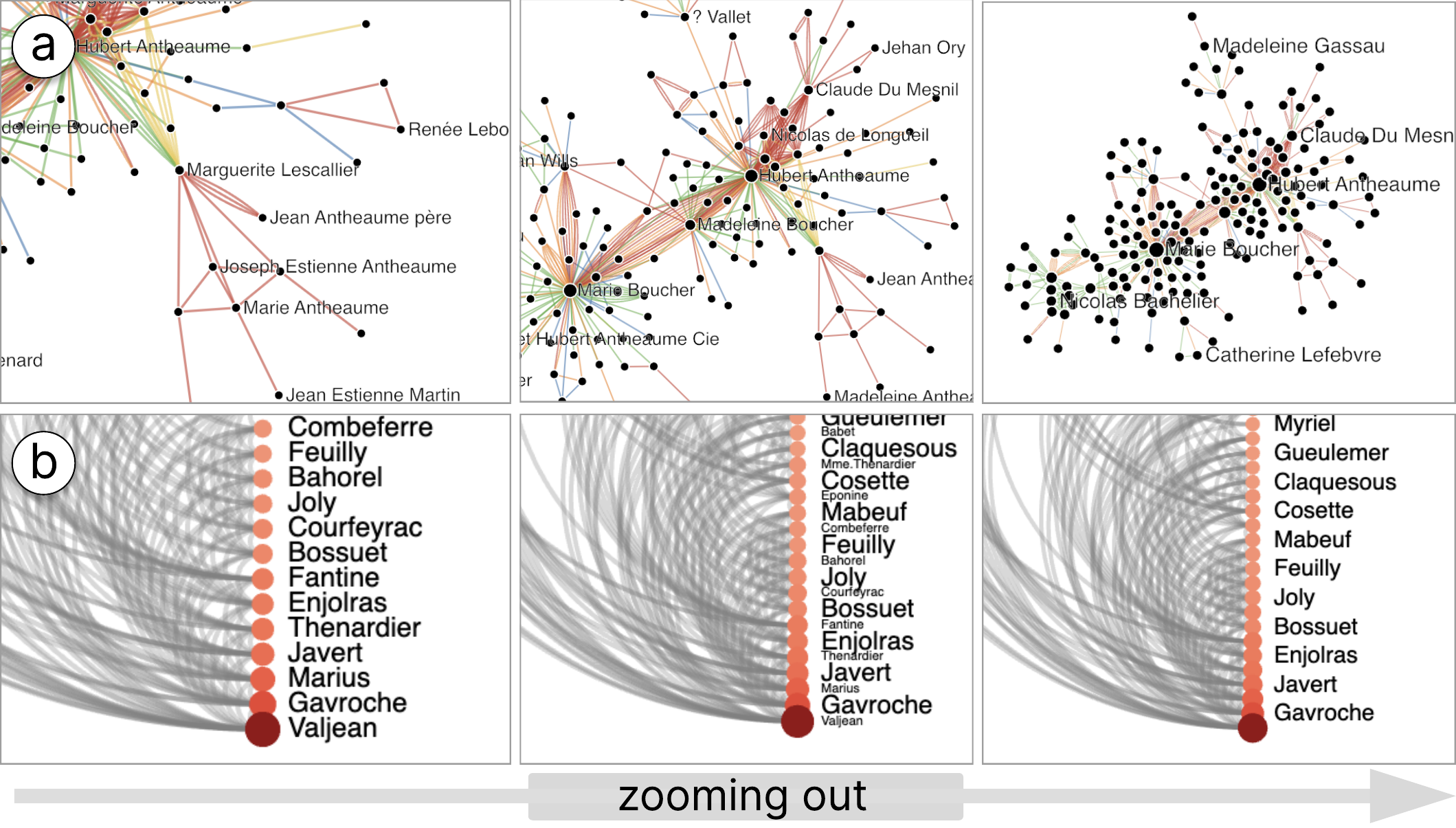}
    \caption{Two examples of gradual label removal to avoid overlap. From left-to-right: zooming out, some labels are first shrunk, then removed. 
  }
    \vspace{-1.5em}
    \label{fig:labelOcclusion}
\end{figure}

In its most simple case, labels can be shown based on a local importance function, e.g., for specific node attributes such as node types or certain degree thresholds. For importance based on a node's local \textit{context}, \grammar{} implements a label removal strategy (\codeowna{overlapRemoval}{true}) inspired by label visibility on geographic maps (shown in \autoref{fig:labelOcclusion} at different zoom levels).
It first calculates the overlap between each two label bounding boxes, then removes the label with the lower attribute value specified by a \codeown{field} or \codeown{expression} parameter. For example, we can specify that the node with the higher node degree or a specific node type should remain visible. In case of a match, a random label is chosen. The distance at which two labels qualify as overlapping is set by the \codeowna{overlapDistance}{[0,infinite]}. A value of \texttt{1} means that label bounding boxes can touch to trigger an overlap; a value of \texttt{0} will allow node labels to fully overlap (no removal of labels); a value of, e.g., \texttt{0.5} will allow 50\% label overlap while a value of \texttt{1.5} means that overlap is defined at less then 50\% white space between the two bounding boxes.
If this overlap occlusion strategy is active, it automatically updates label visibility upon each zoom or other interaction that might change label visibility (\autoref{fig:labelOcclusion}). 

To avoid disruption from labels suddenly appearing and disappearing when zooming, a developer can specify a \codeowna{shrink}{true} parameter so that labels are scaled by half their size when an overlap is detected, before being hidden (\autoref{fig:labelOcclusion}(b)). Beyond force-directed networks, these mechanisms can help render \textit{any} type of layout and visual representation more readable.

\subsection{Interaction}
\label{sec:interaction}

Interaction for network visualization is notoriously tricky because it often affects the layout and elements' visual encoding. We differentiate between two types of interactions: interaction for exploration and interaction for styling. 

\subsubsection{Interaction for exploration}

\begin{figure}[h]
    \centering
    \includegraphics[width=1\columnwidth]{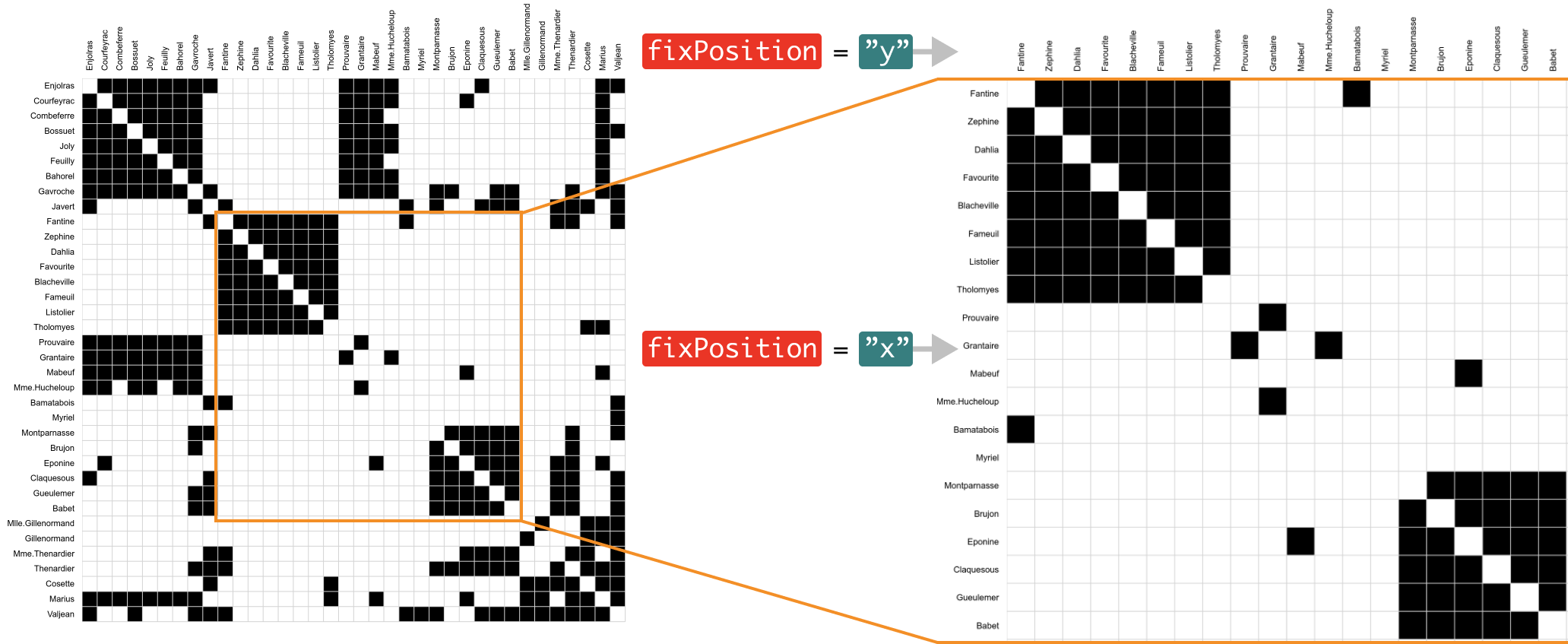}
    \caption{Example of fixing labels to screen positions when panning and zooming a visualization using \codeown{fixPosition}. 
     The labels' size is constant by default, while the cells' size is scaled with the zoom using \codeowna{scaling}{true} in the mark definition.
    }
    \vspace{-1em}
    \label{fig:zoom}
\end{figure}

Interactions such as mouse-over tool-tips, re-positioning nodes by dragging, and other mouse-over behaviors are specified in the definition of the corresponding \codeown{mark} in the \codeown{vis} block. A global option determines whether panning and zooming are enabled or disabled within a specification to allow, e.g., maintaining the view on a defined portion of the network in a storytelling context while allowing for other interactions to be active. Interactions for tooltips, node dragging, and mouse-over behaviors are specific to individual visual marks. 
The effect of zooming on marks' sizes can be controlled with the \codeowna{scaling}{(true|false)} property. If false, marks are not scaled during zoom and always have the same constant size by transforming layout coordinates rather than screen coordinates (see \autoref{fig:zoom}). 

However, default panning and zooming, even if elements maintain their sizes, can remove important elements from the screen, such as labels in matrices and other linear and Cartesian layouts. In these cases, \grammar{} allows the movement of visual marks during panning and zooming to be restricted in either or both screen dimensions: \codeowna{fixPosition}{("x"|"y"|"both"|none)}. For instance, declaring \codeowna{fixPosition}{"x"} on the row labels of an adjacency matrix (or any other Cartesian or linear layout), fixes those labels at their initial x-screen-position, effectively preventing them from exiting the screen whilst panning and zooming the cells inside the matrix (see \autoref{fig:zoom} for an example with matrix).

Another use case for interaction is allowing specific nodes to be highlighted whilst others are de-emphasized. A \textbf{selection interaction} can update a set of nodes and links defined in a \codeown{selection}. For example, clicking a node or selecting nodes and links with a lasso interaction can add, remove, or toggle membership of the corresponding nodes or links in a selection. The visual appearances for these nodes and links can then be updated, such as by changing their color. 

\textbf{Selections} can also be defined computationally through topological transformations (\autoref{sec:transformations}), metrics (\autoref{sec:metrics}), or groupings.
For example, a specification may define an interaction that sets one selection to contain the node that was most recently clicked on, and a second selection could be defined to contain the contents of the first selection plus any surrounding nodes; or highlight nodes of different timesteps. The nodes in those selections can then be styled accordingly.

\subsubsection{Interaction for Styling and highlighting}

\begin{figure}[h!]
    \vspace{-.5em}
    \centering
    \includegraphics[width=1\columnwidth]{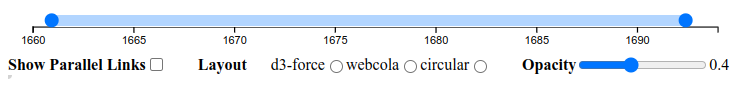}
    \caption{
    \grammar{} can bind parameters to interactive controls.
    }
    \vspace{-1.5em}
    \label{fig:parameters}
\end{figure}

Interactive styling allows the designer to enable specific personalizations by users; 
for example, choosing from a set of node-ordering options, changing layout parameters, or adjusting the size of nodes and links. \grammar{} provides \codeown{parameters} as global public variables that  can be bound to UI input elements such as sliders (for time or other attributes) or select boxes. Interacting with these UI elements updates the values of the bound parameters in the visualization; 
the parameters can be used in constructs such as \codeown{layouts}, \codeown{orderings}, or \codeown{vis} (where they may affect transparency, color, sizes, etc.) (\autoref{fig:parameters}). 
Parameters can also be used in expressions, for example to allow interactive node filtering and create the impression of a temporal flip-book controlled by a time slider (e.g.,~\cite{bach2014visualizing}).

\section{Implementation}
\label{sec:implementation}

We implemented a reference implemention of \grammar{} as a TypeScript library that accepts a specification in JSONC format (JSON with Comments), loads the required data, and renders the resulting visualization to an SVG or HTML Canvas element. This library is available as the \texttt{netpanorama} package on \texttt{npm}.
We also implemented an interactive editor, which lets a user enter a specification and view the resulting visualization side-by-side.
Accepting JSONC not only allows the inclusion of explanatory comments, but also allows users to comment-out sections of a specification as they experiment with alternatives. Internally, our implementation creates its own data structures for tables and networks. 

\grammar{} uses several libraries that are part of the Vega project, for 
fetching and parsing data from a CSV/JSON file (\texttt{vega-loader}), generating a D3~\cite{D3} scale function from a JSON specification (\texttt{vega-encode}), parsing and evaluating expressions in a safe subset of JavaScript (\texttt{vega-expression}), and actually rendering a set of visual marks within an SVG or Canvas element (\texttt{vega-scene\-graph}). However, our \grammar{} implementation is a separate system that re-uses some components of the Vega system, rather than being an extension of Vega. For topological layout computation \grammar{} delegates to D3.layout, WebCoLa/SetCoLa, Tulip~\cite{tulip5}, and Cytoscape.js. Most of these libraries run in the browser with the exception of Tulip, a C++ framework, for which we created a Python wrapper that provides an HTTP API: networks are sent by a POST request to our server, and the node positions are returned. Matrix seriation is performed using \texttt{reorder.js}~\cite{reorderJS}, the only suitable JavaScript library we are aware of.

Notably, the current system of reactivity and interaction---even if inspired by Vega---is specific to \grammar. 
\grammar{} models the relationships between the different items of a specification with a dependency graph. When an item of the graph is re-evaluated (for example, following an interaction or a change in a parameter bound to an input), the depending graph items are also recomputed.
For instance, if the size of a visual mark construct depends on a \codeown{nodeSize} parameter, then changing this parameter value using an input will cause the re-evaluation of the mark specification and recompute the scenegraph values. However, marks not depending on this parameter would not be recomputed, enabling fast interactive visualizations.
Some dependencies are also hard-coded; for example, if a text mark is specified with a label overlapping removal strategy, any zoom action will trigger a re-computation of the text marks as the zoom level affects which text marks are shown and which are not. 
\section{Examples}
\label{sec:examples}

Our main goal with \grammar{} is to provide a unified and efficient way to declaratively describe a wide range of network visualizations beyond node-link diagrams (C2), for multivariate, temporal, and geographic networks (C1), including all necessary steps from data loading via transformations and analysis (C3) to visualization and interaction (C5), all while building on existing toolkits and libraries (C4).
In this section, we demonstrate how \grammar{} satisfies these criteria across two scenarios: 
\textit{a)} re-creating existing network visualization techniques from the literature, and
\textit{b)} building bespoke applications that embed \grammar{} visualizations.

\subsection{Recreating existing visualization designs}
\label{sec:existingdesigns}

Using \grammar{},
we are able to create many common network visualization techniques, as listed below.
Some of them are shown in \autoref{fig:teaser} with more on our website. We are not aware of any other toolkit providing this many non-standard designs.

\begin{itemize}[noitemsep,leftmargin=*]
    \item 
\textbf{Dynamic networks:}
time-glyphs in matrices~\cite{brandes2011asymmetric}, 
small multiples for nodelink diagrams~\cite{bach2014visualizing}, 
a biofabric layout visualization~\cite{william2012combing}, 
animated nodelink diagrams with timesliders~\cite{baur2001visone,Vistorian}, 
parallel edge splatting~\cite{burch2011parallel}, time-arcs~\cite{TimeRadarTrees},
linkwave~\cite{riche2014linkwave}.

\item 
\textbf{Multivariate networks:}
Pivot Graphs~\cite{wattenberg2006visual}, 
Jigsaw~\cite{stasko2008jigsaw}, 
GraphDice scatterplots~\cite{bezerianos2010graphdice},
Semantic substrates~\cite{shneiderman2006network},
hive plots~\cite{krzywinski2012hive},
matrix cell glyphs for multivariate links~\cite{vogogias2020visual}, link weights~\cite{chang2017evaluating}, and link encodings in nodelink diagrams~\cite{Holten_2011}, wiggly lines from Abyss-explorer~\cite{abyss_explorer},
bimatrices

\item
\textbf{Modifications of nodelink-diagrams}:
arc-diagrams~\cite{wattenberg2002arc}, 
circular and concentric node-link diagrams, and layered graphs

\item
\textbf{Dense networks}: 
adjacency lists, 
adjacency matrices~\cite{henry2006matrixexplorer} with diverse reordering methods~\cite{behrisch2016matrix}, node grouping~\cite{ontotrix}, and glyph encodings for comparisons~\cite{alper2013weighted,vogogias2020visual} and time~\cite{yi2010timematrix,stein2010pixel}, also matrix small multiples~\cite{perer2012matrixflow}.

\item\textbf{Hybrid techniques:} NodeTrix visualization~\cite{henry2007nodetrix}, MatLink~\cite{henry2007matlink}, circular diagrams on maps, node-link diagrams inside cluster nodes.

\end{itemize}

\subsection{Application use cases}
\label{sec:applications}

We used \grammar{} to specify network visualizations for a set of projects that required bespoke designs in specific applications and contexts. Working on these projects whilst developing \grammar{} helped to inform many of the features in \grammar{} and to improve its usability.
It motivated the inclusion of specific functionality including strategies for handling overlapping node labels and fixing the position of labels during panning and zooming (\autoref{sec:labeling}), interactive parameters for styling (\autoref{sec:interaction}), as well as the general range of visualization designs \grammar{} should be capable of expressing.

For an online platform for visualizing multivariate, temporal, and geographic networks, we build 11 interactive visualizations from the list in \autoref{sec:existingdesigns}. Each visualization is highly interactive with tooltips, highlighting neighbors, pan, zoom, label occlusion strategies, a time-slider (for temporal data) and diverse visual styling parameters. All visualizations have parameterizable visual attributes and a timeline to filter nodes and links by time (\autoref{fig:parameters}). Some visualizations, such as matrices, arc-diagrams, and circular diagram can chose among a set of orderings and seriations. By default, label visibility is determined by our occlusion mechanism (\autoref{sec:labeling}) while labels in matrices, arc-diagrams, timelines and adjacency lists are indifferent to pan and zoom (\autoref{sec:interaction}). All visualizations are synchronized through brushing and linking and we made sure visual encodings (link weight, link direction, node size, color schemes) are consistent across all visualizations, which took significant iterations to the designs. Internally, that platform generates the \ca{data} and \ca{network} constructs 
based on the user's inputs, and uses these to complete a template \grammar{} specification.

The three other applications we built include an analytical web-application for literature analysis featuring three visualizations of co-authorship networks; a storytelling website explaining social relationships among intellectuals; and an overview dashboard showing links between peace agreements and the associated countries.
In all scenarios, we prototyped and reused visualizations with \grammar{} and iterated on them with our project partners. 
Each application was then optimized for its respective tasks by adding or removing interactivity, creating specific network models, calculating specific metrics, and adapting the visual appearance to the given context and audience. 
Specifically, no context switching nor
additional libraries were needed during the whole development pipeline from data processing to the visual encoding choices since every step was done inside \grammar{} using a consistent grammar. 
Moreover, each step of the development and design process was potentially reusable for other designs and features in the other projects and \grammar. For instance, once we designed an arc-diagram, it was very easy to create a new specification for a matrix visualization, using the same network, metrics, scales, and orderings. Customization was also easy using the low-level properties of the \codeown{vis} construct---such as the \codeown{fontsize}, \codeown{stroke}, and \codeown{fill} properties---and easily reusable across several design of the same dashboard when design consistency was needed.

Across all those use cases, \grammar{} has sped up our design process by shifting our energy from implementation to discussing design requirements and complementary views for each application: 
\textit{What's a good ratio of node to edge size in a visualization? 
Which orderings make most sense for a given data set and audience? 
What is the minimal set of visualization designs required to solve the tasks?
Are those designs suitable for data sets with different characteristics (e.g., density, number of node types), 
Can we combine ideas from different visualization designs? How can we improve current designs?  
} and 
\textit{How do we balance the distribution of information across multiple visualization?} In hindsight, we have found ourselves following a ``more-is-more approach'' for each of our user-centered co-design projects: building lots of simple visualizations first, triaging by usefulness, then aggregating and complementing.
Besides speeding up our design process, 
\grammar{} helped us show unfamiliar designs to our collaborators ready with their data, 
instead of discussing abstract sketches and examples: we could show what information we can show about their data and discuss the actual usefulness of this information---and in consequence the designs. 
Finally, \grammar{} has given our teams a language to talk to our collaborators and hand over the designs for each visualization. As another result, we now have a wide range of ready-to-use \grammar-templates 
that can be adapted to different data formats, network schemas, visual design requirements, and interaction needs.

\section{Discussion}
\label{sec:discussion}

\grammar's expressiveness extends that of existing network visualization toolkits, while integrating many existing concepts. It further aims to help reify
common concepts and design patterns for the visualization of relational data. By defining declarative constructs for ordering, network construction, network transformation, layouts, labeling, visual marks, and interaction, we identify those ideas and help translate concepts from one technique or design to another. We hope \grammar{} will help to further unify those ideas and accelerate the application and dissemination of efficient 
network visualization techniques through a unifying medium: visualization designs can be shared online and adapted for specific applications on demand while providing crucial transparency about network transformation (e.g., filtering) and visualization (e.g., layout) if published alongside the visualizations. We also hope that \grammar{} and more formal ways of describing visualization designs can help with the evaluation of the respective designs and underlying algorithms, e.g., changing one design parameter at a time and avoiding (re)implementation of many designs. \grammar{} could help with the education in network visualization by defining concepts and allowing easy prototyping and design space exploration~\cite{bach2023challenges}. 

\textbf{Extensions}---We believe no grammar will ever be likely to capture the entire set of design possibilities~\cite{mcnutt2022no} but we see a range of extensions or purpose-specific simplifications to \grammar, such as higher-level templates for common designs such as matrices, arc diagrams, maps, etc.
For this first version of \grammar, we excluded a range of concepts required for some visualization techniques 
such as 3-dimensional techniques (e.g., ~\cite{brandes2003visualizing,bach2014visualizing}), which would likely require further interaction techniques, such as unfolding a matrix cube~\cite{bach2014visualizing} or applying a 3D distortion~\cite{carpendale1997extending}. 
Likewise, we did not explicitly consider trees~\cite{GoTree}, set structures~\cite{STAR_groups}, or hypergraphs~\cite{buono2021hypergraph}.

Intermediate network structures such as ego-networks, graphlets, and subgraph patterns---e.g., cliques---could also be added to \grammar{} in the future. However, These concepts do not fit the current \codeown{transform} abstraction and may need a new type of construct, or could be specified within the \codeown{grouping} capabilities. Those network structures could be used to render other visualization techniques, including motif simplification~\cite{dunne2013motif}. 

Temporal networks could be supported further through advanced animation and transition techniques~\cite{bach2014visualizing}, ideally integrating existing grammars (Gemini~\cite{Gemini}) as well as stabilized layouts~\cite{archambault2012mental}. Finally, while \grammar{} can render node-link diagrams on maps, retrieve coordinates for location names, and can individually arrange nodes that share the same geographic locations (\autoref{fig:teaser}), extending \grammar's expressiveness to more bespoke visualization techniques for geographic network visualizations~\cite{geospatial_networks}, will require dedicated handling of geographic data, distortion, and aggregation.

Some current limitations in the expressiveness of \grammar{} are due to its semantics specifying certain things individually, rather than jointly.  For example, \grammar{} provides options to control edge shape but currently processes each edge shape individually rather than being able to jointly adjust all edges for edge bundling~\cite{STAR_bundling}. The same applies to temporal stabilized layouts, already mentioned above~\cite{archambault2012mental}. Adjacency matrix seriation could also be performed jointly for several matrices~\cite{simultaneous_ordering}.
More
specialized layouts or orderings (such as  metro-lines, confluence graphs~\cite{bach2016towards}, multi-dimensional scaling~\cite{bach2015time,van2015reducing} or the custom re-ordering technique in Massive Sequence Views~\cite{massive_sequence_views}) could be implemented and accessed as specific layouts.

Advanced interaction techniques for aggregation and exploration (e.g.,~\cite{interactive_curvature,bach2015small}) 
could help tame larger and more complex networks and modify a network at different levels: its network structure, its layout, and visual mappings (e.g., semantic zoom, or portals~\cite{hadlak2011situ,ghani2011dynamic}, areas~\cite{interactive_curvature}, link-sliding~\cite{moscovich2009topology}, or interactive link bundling.
Care is needed to integrate such interactions in a conceptually unified way, rather than as an \textit{ad hoc} grab-bag of isolated functionality. Performance and scalability to large graphs beyond aggregation and providing basic interactivity was not initially a priority; we imagine implementations in WebGL.

\textbf{Future Tools}---We think \grammar{} could lay the foundations for an ecosystem of tools, including more concise languages and templates, and graphical interfaces that augment the textual specifications~\cite{construction-interface-survey}. We imagine \grammar{} being used in web applications with network visualization components in analytical scenarios as well as data-driven storytelling where these visualizations are embedded into other comics and further adapted to the format of choice (e.g.,~\cite{bach2016telling,wang2021interactive}). 
Given our many example specifications, recommender systems could propose visual representations for a particular dataset or task, e.g., by learning from a user's previous designs. We can also imagine higher-level specifications aimed at building entire network applications that, besides referencing visualization \grammar{} visualization designs, include concepts for annotation, multiple view layout, responsiveness, and accessibility.

\bibliographystyle{abbrv-doi-hyperref}

\bibliography{refs}

\end{document}